\documentclass[10pt, a4paper]{article}
\usepackage{lrec2022} 
\usepackage{multibib}
\newcites{languageresource}{Language Resources}
\usepackage{graphicx}
\usepackage{hyperref}
\usepackage{tabularx}
\usepackage{soul}
\usepackage{titlesec}
\titleformat{\section}{\normalfont\large\bfseries\center}{\thesection.}{1em}{}
\titleformat{\subsection}{\normalfont\SmallTitleFont\bfseries\raggedright}{\thesubsection.}{1em}{}
\titleformat{\subsubsection}{\normalfont\normalsize\bfseries\raggedright}{\thesubsubsection.}{1em}{}
\renewcommand\thesection{\arabic{section}}
\renewcommand\thesubsection{\thesection.\arabic{subsection}}
\renewcommand\thesubsubsection{\thesubsection.\arabic{subsubsection}}

\usepackage{epstopdf}
\usepackage[utf8]{inputenc}
\usepackage{amsmath}

\usepackage{hyperref}
\usepackage{xstring}

\usepackage{color}

\title{\textbf{A Dataset for Speech Emotion Recognition in Greek Theatrical Plays}}

\name{Maria Moutti$^1$, Sofia Eleftheriou$^2$, Panagiotis Koromilas$^2$, Theodoros Giannakopoulos$^2$} 

\address{$^1$University of the Peloponnese, $^2$National Center for Scientific Research - Demokritos \\
         \{mar.moutti\}@gmail.com, \{seleftheriou, pakoromilas, tyianak\}@iit.demokritos.gr\\}

\abstract{
Machine learning methodologies can be adopted in cultural applications and propose new ways to distribute or even present the cultural content to the public. For instance, speech analytics can be adopted to automatically generate subtitles in theatrical plays, in order to (among other purposes) help people with hearing loss. Apart from a typical speech-to-text transcription with Automatic Speech Recognition (ASR), Speech Emotion Recognition (SER) can be used to automatically predict the underlying emotional content of speech dialogues in theatrical plays, and thus to provide a deeper understanding \emph{how} the actors utter their lines. However, real-world datasets from theatrical plays are not available in the literature. 
In this work we present \emph{GreThE}, the Greek Theatrical Emotion dataset, a new publicly available data collection for speech emotion recognition in Greek theatrical plays. The dataset contains utterances from various actors and plays, along with respective valence and arousal annotations. Towards this end, multiple annotators have been asked to provide
their input for each speech recording and inter-annotator agreement is taken into account in
the final ground truth generation. In addition, we discuss the results of some indicative experiments that have been conducted with machine and deep learning frameworks, using the dataset, along with some widely used databases in the field of speech emotion recognition. 
\\ \newline  }

\begin{document}

\maketitleabstract

\section{Introduction}
\label{sec:intro}

The task of recognizing the underlying emotion from speech, irrespective of its semantic content, is rather important in various applications. However, it is hard to notate even by human beings, mostly due to the subjectiveness of the emotional content. The ability to automatically conduct it, is a demanding task and still an ongoing subject of research. Several open-source databases exist in the field of Speech Emotion Recognition (SER) which may contain audio-only or multimodal information and are usually annotated on two categories: categorical attributes (distinct classes of emotions) \cite{psycology2,Ekman} and dimensional attributes (continuous values of valence, arousal and intensity) \cite{wundt2011outlines,circumplex}. The emerge of the fields of Machine and Deep Learning in the past decade, gave the chance to researchers to apply research outputs on real-world problems. However, despite the fact that SER is a task that has gained great attention in the literature \cite{koromilas2021deep}, the industrial applications of the proposed works are either centered around web content (online video or podcast analysis) or have been applied on actual conversations to enrich understanding (eg. empathetic dialogue \cite{ma2020survey}). 

At the same time, cultural events are a base factor for every human civilization and, as such, they can also benefit from modern Machine Learning (ML) applications. Automatic review mining, automatic summary generation of movies, content-based movie recommendation systems, music and movie retrieval, production of subtitles/transcriptions in guided tours, movies or theatrical performances are just some examples. These applications of ML on cultural content changes the way the content is generated and distributed. Moreover, it provides solutions to increase \emph{inclusion} of particular groups of the population: e.g. automatic generation of subtitles, enriched with paralinguistic attributes such as emotional arousal could help people with hearing loss to actually understand and "feel" a theatrical play. 

The widely used SER approaches have not properly been used to address challenges that are set from cultural content data, with the best example being the theatrical plays.

This is mostly due to the fact that methods that are trained on SER datasets cannot be properly applied on theatrical content. This claim is based on the following facts:
\begin{enumerate}
    \item actors that perform in theatrical plays are more expressive and thus the emotional levels are aroused. That is, the arousal classes are shifted towards more energetic emotions (e.g. the weak class is expressed in a more "aroused" - energetic way). As far as the valence classes are concerned, they also differ to the respective valence classes in other SER datasets. For example, depending on play type (e.g. dramas) the neutral class itself can also include negative or positive emotions, compared to SER datasets that try to capture a real-world (non-theatrical) context. 
    \item interaction with the audience make the actors express their actions in different ways so as to be better perceived by the attendants. That is, actors use emotional states that are not common in real-life conversations and thus are not included in the general SER datasets
    \item the recording setups and conditions of theatrical plays differ from that of the datasets found in the literature, as the former may include complex microphone systems and fine post-editing procedures. 
\end{enumerate}

In this work, we propose the Greek Theatrical Emotion (GreThE) dataset with the aim of filling the existing gap in the literature. GreThE is a collection of speech utterances from 23 Greek theatrical plays annotated with regards to the respective levels of emotional valence and arousal. We also provide a baseline evaluation for the presented dataset and we examine whether the domain knowledge of general SER can be used to achieve robust performance on GreThE.  

The paper is organized as follows:
in section \ref{sec:related} we report the related works on SER datasets, section \ref{sec:dataset} describes the GreThE dataset, section \ref{sec:methods} contains the used classification methods, section \ref{sec:experiments} reports the experimental results, section \ref{sec:availability} comment on the availability of the dataset and section \ref{sec:conclusion} concludes the paper.

\section{Related Work} 
\label{sec:related}

\subsection{Speech emotion recognition datasets}\label{ssec:ser_data}
Remarkable effort has been given in the literature, to create emotion-based datasets that accurately represent the basic human emotions and reactions in speech signals. The existing datasets can be classified into four categories according to the recording procedure that is followed through the data collection process \cite{koromilas2021deep}. Specifically, these methods include one of the following: \textit{(i)} \emph{spontaneous} speech: the participants are unaware of the recording while their speech and reactions are recorded with hidden mechanisms in a real environment \cite{cao2015speaker}; \textit{(ii)} \emph{acted} speech: the emotional condition of the speakers is acted; \textit{(iii)} \emph{elicited} speech: where the speaker is placed in a situation which evokes a specific
emotional state \cite{basu2017review}; and \textit{(iv)} \emph{annotated public} speech: data from public sources, such as YouTube, are annotated to associate them with a range of emotional states.

Some of the most commonly used datasets in that field are: \emph{IEMOCAP} \cite{iemocap}, a multimodal database which includes recordings from 10 actors annotated in categorical and dimensional attributes, \emph{Emo-DB} \cite{emo-db}, an emotional speech database containing recordings of 10 speakers that simulates 7 emotional states, \emph{MSP-podcast} \cite{podcast} that contains speech segments from podcast recordings which are annotated with emotional labels using attribute-based descriptors and categorical labels, \emph{EMOVO} \cite{emovo} an Italian emotional speech database created by 6 actors that simulates 7 emotional state, \emph{SAVEE} \cite{savee} database which describes the emotion in 6 distinct categories and \emph{RAVDESS} \cite{ravdess} that provides speeches of 24 actors and songs in audio and video format and includes 7 emotional expressions among with two levels of emotional intensity. 
 
\subsection{Greek emotion recognition}
The respective work on Greek-based speech emotion recognition databases is limited. In particular, one of the first approaches has been introduced in the \emph{AESDD} dataset \cite{aesdd}, which is a publicly available SER database that contains utterances of acted emotional speech in the Greek language created by 5 actors and annotated with five emotional states (without containing the neutral state). Furthermore, \emph{SEWA} \cite{sewa} is a multi-lingual database for audio-visual emotion and sentiment research in the wild containing more than 2000 minutes of data of 398 people coming from 6 cultures (including Greek), annotated among others in terms of continuously valued valence and arousal. 

\subsection{Emotion datasets for cultural content}
As discussed in \ref{ssec:ser_data},  the participation of actors in the recording of emotional databases in order to perform acted emotional speech has been a popular approach in the study of emotions. There is a range of databases that are based on actors, such as CREMA-D \cite{crema-d}, CaFE \cite{cafe}, IEMOCAP \cite{iemocap}, EMOVO \cite{emovo} and RAVDESS \cite{ravdess}. 

The MSP-IMPROV corpus \cite{msp-improv} is an example of the elicited speech (category \textit{(iii)}) datasets. It proposes an alternative, approach according to which the authors define hypothetical scenarios for each sentence that are carefully designed to elicit a particular emotion. Two actors improvise these emotion-specific situations, leading them to utter contextualized, non-read renditions of sentences that have fixed lexical content and convey different emotions. In this way, they manage to produce more natural behaviors. However, neither the recording conditions or the emotional reactions can be considered to be close to these of an actual theatrical play which is more expressive and differs from real-life reactions.

Regarding the task of recognizing emotions in actual theatrical plays, to our knowledge, the only study in the literature is presented in \cite{Gloor2019MeasuringAA}, where the authors developed a system to measure both audience and actor satisfaction during a public performance. They used smartwatches to gather physiological signals from the actors, as well as video cameras to capture facial expressions from the audience and finally speech signals from the actors to be used in SER. Then, predictions of emotions were extracted on the three channels of information using pretrained models from existing external datasets and they presented results of correlation metrics between the emotions predicted from the individual channels. Therefore, the particular work does not present a new speech emotion recognition dataset rather than it examines relationships between \emph{predicted} emotions of both the audience and the actors from different channels of information. 

Cinematic films is a type of content with limited representation in the literature of emotion recognition. Specifically, the EMOVIE \cite{emovie} dataset that includes 9,724 samples from seven movies with audio files annotated in the emotion polarity and the AVE \cite{ave} dataset which is based on an Indonesian Movie study \cite{indonesian} are two of the few examples.

Our proposed dataset, GreThE, aims to fulfill the gap in the literature of language resources, by proposing a publicly available non-english speech emotion recognition dataset for real-world theatrical recording conditions.

\section{Dataset}
\label{sec:dataset}
\subsection{Audio data collection} \label{ssec:datacollection}

We have selected to adopt the \texttt{Audacity} open source audio editing tool to  manually segment at least 20 single speaker utterances from each theatrical play. This process led to 95 single-speaker utterances (90 unique speakers) from 23 Greek discrete theatrical plays, resulted to a total of 500 recordings/speeches. The total duration of speech is 46 minutes and their average duration is 5.5 seconds (the shortest utterance is 2.1 seconds and the longest utterance is 10.9 seconds).

\subsection{Utterance annotation process} \label{ssec:annotation}
Each speech utterance that has been collected from the various theatrical plays, as described in Section \ref{ssec:datacollection}, has been annotated with respect to its emotional content. Towards this end, we have selected to use the dimensional emotional representation of Valence and Arousal. Our goal was to adopt the standard in SER 3-class approach \cite{metallinou2012context}, according to which the classes for Valence are: negative, neutral and positive and the classes for Arousal are weak, neutral and strong. 

However, in the initial annotation process we asked the individual annotators to provide their feedback in a 5-valued scale for both tasks. Then we used aggregates on these 5-scale estimates to map them to the three distinct final ground truth classes as described in Section \ref{ssec:annotationaggregation}. So in the individual annotation procedure, the following 5 labels were used: (1) very weak (2) weak (3) neutral (4) strong (5) very strong, and (1) very negative (2) negative (3) neutral (4) positive (5) very positive, for the arousal and valence task respectively.

The annotation process has been carried out by four individuals. We have selected to adopt the \texttt{Label Studio} open source data labeling tool, that provides a web-enabled dynamic graphical interface for annotating multimodal content. \footnote{\url{https://labelstud.io}}

\begin{figure}[!h]
\begin{center}
    \includegraphics[width=\linewidth]{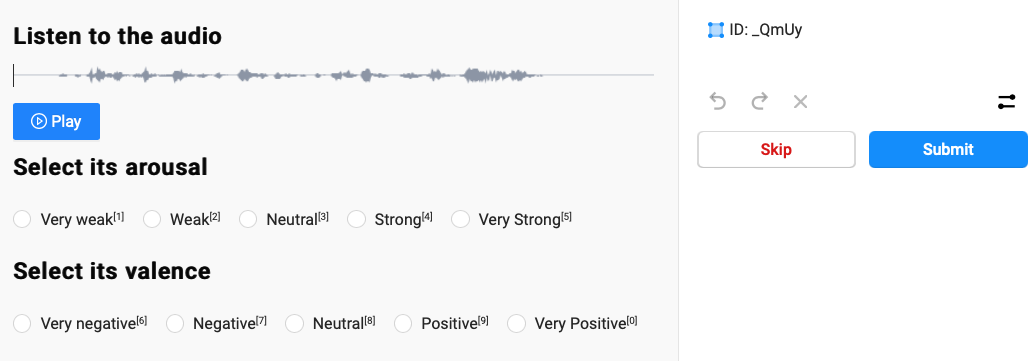} 
    \caption{Screen shot from the LabelStudio project used for annotating the utterances.}
    \label{fig:label_studio}
\end{center}
\end{figure}

Each individual annotated the whole dataset of 500 utterances and the final ground truth has been generated by an annotation aggregation procedure described in Section \ref{ssec:annotationaggregation}.

\subsection{Annotations aggregation} \label{ssec:annotationaggregation}

After the previously described data annotation process, we applied an aggregation step to extract the final 3-class ground-truth label for each data point (utterance) from the initial individual 5-scale annotations. Furthermore, for each data point we calculated the agreement between the individual annotators so as to quantify the level of complexity of the adopted classification tasks (valence and arousal) for the particular dataset \cite{icnlsp}. 

Final ground truths were obtained by first calculating an average annotation rating (in the 1-5 scale for both classification tasks, as described above) and comparing it against some predefined thresholds, so as to conclude a final 3-class classification for the respective sample. In particular, two thresholds are required to map the aggregated annotations to a 3-class taxonomy. An obvious selection would be to uniformly select the thresholds in the 1-5 range: in that case the thresholds would be $T1=2.33$ and $T2=3.66$, i.e. any utterance with an average annotation in the $[T1, T2]$ range would be finally assigned to class "neutral". 

However, we have made some slight modifications from that uniform selection of thresholds, to satisfy the aforementioned (Section \ref{sec:intro}) shift of the theatrical utterances to more aroused emotional states. For the valence task, the $T1$ was defined to be equal to 2.33 and $T2$ equal to 3.33.  Concerning the task of arousal, we used slightly different thresholds: the lower threshold was defined to 2.66 and the higher threshold to 3.66. 
This minor differentiation between the thresholds for the two tasks stems from the fact that the theatrical data are characterized by a high dominance of negative and strong emotions. So, for instance, the fact that "weak" arousal would be rather rare, setting the lower threshold $T1=2.66$ instead of the "default" uniform value (2.33) makes it possible to decide in favor of the weak class even in borderline cases such as the following: suppose the 4 annotators have given the ratings "weak" (value 2), "weak" (value 2), "neutral" (value 3), "neutral" (value 3), the average value of these ratings is $\frac{2 + 2 + 3 + 3}{4} = 2.5$. In that case, having the uniform $T1$ would make that sample be finally classified as "neutral", while the threshold $T1=2.66$ finally classifies that sample as "weak".

To sum up, the allocation of the final labels followed the below thresholding rules: 
\begin{equation*}
  arousal_i=\begin{cases}
    \text{weak}, & E[A_{i}] \leq 2.66.\\
    \text{neutral}, & 2.66 < E[A_{i}]< 3.66.\\
    \text{strong}, & E[A_{i}] \geq 3.66
  \end{cases}
\end{equation*}

\begin{equation*}
  valence_i=\begin{cases}
    \text{negative}, & E[V_{i}]\leq 2.33.\\
    \text{neutral}, & 2.33 < E[V_{i}] < 3.33.\\
    \text{positive}, & E[V_{i}] \geq 3.33
  \end{cases}
\end{equation*}

where $A_{i,\alpha}$ is the arousal annotated value for a sample $i$ of the annotator $\alpha$ (in the 1-5 range) and $E[A_{i}]$ is the mean annotation of sample $i$ (similarly for valence, let $V_{i,\alpha}$ be the annotation value for valence in the 1-5 range)

In addition to the mean values of valence and arousal annotations that has been used to extract the final ground truths, through the thresholding steps described above, a deviation threshold is also considered, in order to filter out controversial annotations. To this end, the mean absolute deviation ($MAD$) of the annotations of each sample is calculated for each annotation $X$ for both tasks: \begin{equation*}MAD_i=E(|X_{i,\alpha} - E(X_{i,\alpha})|)
\end{equation*}
This is obviously a metric of inter-annotator agreement for the given sample, therefore samples with high $MAD$ values should be excluded from the dataset. This rule is used as a safety net step to exclude possible examples with extreme inter-annotator disagreements.

\paragraph*{Inter-annotator agreement:}
Apart from using $MAD$ to filter out possible highly questionable utterances, we also used it to compute the overall inter-annotator (dis)agreement for each of the classification tasks. High values of average disagreement would have indicated that annotators had different points of view on the ratings.

On top of the total disagreement for all data, the average disagreement for each annotator was also calculated, to evaluate each human annotator compared to the overall ground truth. To this end, we first calculated the disagreement of the annotation of annotator $a$ for the $i$-th sample. Let $A_{i, a}$ be the value of that annotation (in the 1-5 range), and $M$ be the total number of annotators. Then the deviation of $A_{i, a}$ from the average value of all other annotators for the same sample is:
\begin{equation*}
 u_{i, a}= \left|A_{i, a}- \frac{\sum_{j=1}^M {A_{i, j}}}{M}\right|   
\end{equation*}

The results of the filtering procedure and the calculated agreement metrics for the arousal and valence task are listed in Table \ref{theatrical}. It has to be noted that the average disagreement of both tasks is below 0.5 (0.48 and 0.49 respectively), which is half the size of the neutral class. 

\begin{table*}
    \centering
    \begin{tabular}{|p{2.7cm}|c|c|c||c|c|c|}
    \cline{1-7}
     &\multicolumn{6}{|c|}{\textbf{GreThE}} \\ \cline{2-7}
     &\multicolumn{3}{|c||}{\textbf{Arousal}}&\multicolumn{3}{|c|}{\textbf{Valence}}\\
     \cline{2-7}
     &\textbf{Strong}&\textbf{Neutral}&\textbf{Weak}&\textbf{Positive}&\textbf{Neutral}&\textbf{Negative}\\
     \cline{1-7}
     \textbf{\centerline{\shortstack{Mean \\ Thresholding}}}&$\mu >= 3.66$& $2.6 < \mu< 3.66$ & $\mu<=2.6$ & $\mu>=3.3$ & $2.33 < \mu<3.3$ & $\mu<=2.33$ \\ 
     \cline{1-7}
     \textbf{\shortstack{Number of samples \\ after Mean \\ Thresholding}} &227&180&93&91&218&191\\
     \cline{1-7}
     \textbf{\centerline{\shortstack{Deviation \\ Thresholding}}} & $\sigma<1.3$ & $\sigma<1.3 $ & $\sigma<1.3$ & $\sigma<1.3$ & $\sigma<1.3$ & $\sigma<1.3$\\
     \cline{1-7}
     \textbf{\shortstack{Number of samples \\ after Deviation \\ Thresholding}} &227&180&93&91&218&191\\
     \cline{1-7}
     \textbf{\centerline{\shortstack{Average \\ Disagreement}}} &\multicolumn{3}{|c||}{\textbf{0.48}}&\multicolumn{3}{|c|}{\textbf{0.49}}\\
     \cline{1-7}
    \end{tabular}%
    \caption{Definition of Arousal and Valence Dataset\label{theatrical}} 
\end{table*} 

\section{Baseline Classification Methods}\label{sec:methods}
\subsection{Traditional machine-learning-based approach} \label{ssec:svm}

As a baseline audio classification technique, we have selected to use a set of handcrafted audio features from the time, spectral and cepstral domains, such as Zero Crossing Rate, Spectral Centroid and Mel Frequency Cepstral Coefficients (MFCCs), along with Support Vector Machines as classifiers. In particular, each speech utterance is first split into a sequence of non-overlapping 50 msec short-term windows (frames), and for each frame a set of 68 audio features is computed. At this stage, each speech utterance is represented by a sequence of short-term feature vectors (short-term representation). 

Then, the mean and standard deviation of these features are extracted in a long-term segment of 3 seconds, using 1 second step. According to that, each utterance is represented by a sequence of (68 x 2 = 136) feature  statistics. Finally, we apply a long-term averaging step, which results in a single-vector 168-D representation for the whole utterance (long-term representation). Note that both short-term (matrix) and long-term (vector) representations are provided in the repository of the dataset. 

As a baseline classification method we have experimented with training and evaluating an SVM classifier with an RBF kernel, using the long-term vector representation described above. All respective experiments and feature extraction procedures have been carried out with the pyAudioAnalysis library \cite{giannakopoulos2015pyaudioanalysis}. 

\subsection{Deep-learning-based approach}
\label{ssec:cnn}
Using audio spectrograms (or mel-spectrograms) as image inputs
to Convolutional Neural Networks (CNNs) is a widely adopted approach in the literature of emotion recognition \cite{koromilas2021deep}. In this work we incorporate this methodology as a baseline from the field of Deep Learning. While this method is expected to under-perform when applied to small datasets, it is perfectly suited for testing whether knowledge from large amount of emotion data can be applied to our proposed dataset. For that reason we use the \href{https://github.com/tyiannak/deep_audio_features}{deep\_audio\_features} library to extract mel-spectrograms and train a CNN for our two speech emotion recognition tasks.

\section{Experimental Results}\label{sec:experiments}

\subsection{Experimental Setup}
\subsubsection{Session-independent validation}
The traditional feature extraction and SVM approach described in Section \ref{ssec:svm}, has been evaluated using a repeated random shuffling train/validation split. The "session" ID used to split the data was based on the ID of the theatrical play. In this way, we guarantee that the evaluation results we report in the Results Section are not assuming dependence on the theatrical plays and are therefore realistic. Note, that this ID is provided with the dataset,  as well. Moreover, we have used the aforementioned validation strategy to experiment against different values of the $C$ parameter. Finally, since the datasets is imbalanced in both classification tasks (valence and arousal), we have adopted a data balancing step using either a basic random subsampling (of the dominant classes) or SMOTE oversampling (synthetic minority over-sampling technique, \cite{chawla2002smote}).

\subsubsection{Cross-domain validation}
\label{ssec:cross_domain}
Intending to examine whether the domain knowledge of Speech Emotion Recognition can be transferred on our dataset, we train a CNN in the way described in \ref{ssec:cnn}, on two widely used datasets, the MSP-podcast \citelanguageresource{podcast} and IEMOCAP \citelanguageresource{iemocap}. Following \cite{metallinou2012context} we define two three-class problems on these datasets, namely arousal (or valence) with classes \textit{(i)} weak (or negative) for values in the range [1, 2] (or [1,3]) for the IEMOCAP (or MSP-podcast) dataset, \textit{(ii)} neutral (or neutral) for values in the range (2, 4) (or (3,5]) for the IEMOCAP (or MSP-podcast), and \textit{(iii)} strong (or positive) for values in the range [4, 5] (or (5,7]) for IEMOCAP (or MSP-podcast). 

Three CNNs are trained in total, one for each of the aforementioned datasets and one for their merged combination. The resulted models, namely CNN\_iemocap, CNN\_msp and CNN\_merged, are subsequently used for testing in the GreThE dataset. That is, none of the GreThE instances are used for the CNNs' training, and thus the used models are completely unaware of the recording conditions and acted speech of a theatrical play.   

Our CNN architecture consists of 4 convolutional and 3 linear layers, including batch normalization \cite{10.5555/3045118.3045167} and the LeakyRelu activation function. We trained the model using the Cross-Entropy loss as loss function, while Adam was chosen to be the optimizer with initial learning rate of 0.002 and a reduce-on-plateau learning rate scheduler. 

\subsubsection{Baseline models}

In order to compare our methods with the threshold of a random classifier, one randomized classifier is defined for each of the two approaches. Specifically, since the traditional machine-learning-based approach (section \ref{ssec:svm}) is trained on the GreThE dataset and thus is aware of the sample distribution, it can only be compared to a prior-aware randomized (prior-aware baseline) classifier, ie. a classifier that randomly predicts class $a$ based on the prior probability of class $a$ in the dataset distribution. On the other hand, the cross-domain trained models, as defined in section \ref{ssec:cross_domain}, are not familiar with GreThE and thus their predictions can only be compared with a completely randomized (baseline) classifier.

\subsection{Results}
In table \ref{tab:results} we report the f1 metrics that emerged from our evaluation.
\begin{table}[!h]
\begin{center}
\begin{tabularx}{\columnwidth}{|c|c|c|}
      \hline
      Experiment & Arousal F1 & Valence F1\\
      \hline
      Baseline             & $27\%$ & $26\%$\\
      Prior-aware Baseline & $31\%$ & $30\%$\\
      \hline
      SVM & $53\%$ & $38\%$\\
      SVM - Oversampling & $55\%$ & $40\%$\\
      SVM - Undersampling & $54\%$ & $39\%$\\
      \hline
      CNN\_iemocap & $40\%$ & $37\%$\\
      CNN\_msp & $36\%$ & $34\%$\\
      CNN\_merged & $41\%$ & $34\%$\\
     \hline
\end{tabularx}
\caption{GreThE evaluation results}
\label{tab:results}
 \end{center}
\end{table}

As can be clearly inferred from the provided results, \textit{(i)} the SVM session-independent method achieves 27.9\% relative improvement for arousal and 21.2\% for valence, and \textit{(ii)} the CNN cross-domain method achieves 20.6 \% for arousal and for 17.5\% valence. 

The small, in absolute values, improvement when using cross-domain validation indicates that \textit{(i)} general knowledge from the task of emotion recognition can be transferred to the new domain, but \textit{(ii)} the problem of theatrical speech emotion recognition differs from that of speech emotion recognition. The second point is easily verified by the fact that the actors expressiveness in theatrical plays usually result in increased arousal, shifting (in terms of energy and frequency) the weak and neutral classes towards the strong one.

\section{Dataset Availability}\label{sec:availability}

The dataset is public available at \url{https://github.com/magcil/GreThE}. Each utterance is represented by either (a) a sequence of short-term feature vectors or (b) a spectrogram. Ground-truth is provided in a simple tabular CSV format for both classification tasks (valence and arousal). Finally, the repository also contains Python scripts that demonstrate (a) the aforementioned SVM experimental setup and (b) the adopted feature extraction methods (so that reproducability can also be made possible and combined with other sources of data).

%
%
%
%
%

\section{Conclusion}\label{sec:conclusion}
In this paper we presented the Greek Theatrical Emotion dataset \emph{GreThE}, a new publicly available data collection for speech emotion recognition in Greek theatrical plays. The dataset contains 500 utterances that have been annotated in terms of their emotional content (valence and arousal). Multiple-annotator data have been used to assure annotation quality. To our knowledge, this is the first dataset with real-world speech data from theatrical plays annotated in terms of the underlying emotion. 

Also, we have presented classification performance results for a baseline machine learning classification approach cross-validated on \emph{GreThE}, along with results using \emph{GreThE} as a test dataset for cross-domain deep emotional models, trained on popular datasets of the English language. Results have proven that (a) the task of recognising emotion - and mostly valence - is rather challenging in theatrical data when training from scratch (b) using state-of-the-art datasets from generic SER on cross-language theatrical data is not effective. This indicates that future works in the field of recognizing emotions in theatrical data should probably consider robust domain adaptation techniques using few-shot learning strategies. 

\section{Acknowledgements}

 This research has been co‐financed by the European Regional Development Fund of the
European Union and Greek national funds through the Operational Program Competitiveness,
Entrepreneurship and Innovation, under the call RESEARCH – CREATE – INNOVATE (project code: T2EDK-01359)

\section{Bibliographical References}\label{reference}

\bibliographystyle{lrec2022-bib}
\bibliography{lrec2022-example}

\begin{thebibliography}{}

\bibitem[\protect\citename{Basu \bgroup et al.\egroup }2017]{basu2017review}
Basu, S., Chakraborty, J., Bag, A., and Aftabuddin, M.
\newblock (2017).
\newblock A review on emotion recognition using speech.
\newblock In {\em 2017 International conference on inventive communication and
  computational technologies (ICICCT)}, pages 109--114. IEEE.

\bibitem[\protect\citename{Burkhardt \bgroup et al.\egroup }2005]{emo-db}
Burkhardt, F., Paeschke, A., Rolfes, M., Sendlmeier, W., and Weiss, B.
\newblock (2005).
\newblock A database of german emotional speech.
\newblock volume~5, pages 1517--1520, 09.

\bibitem[\protect\citename{Busso \bgroup et al.\egroup }2008]{iemocap}
Busso, C., Bulut, M., Lee, C.-C., Kazemzadeh, A., Mower~Provost, E., Kim, S.,
  Chang, J., Lee, S., and Narayanan, S.
\newblock (2008).
\newblock Iemocap: Interactive emotional dyadic motion capture database.
\newblock {\em Language Resources and Evaluation}, 42:335--359, 12.

\bibitem[\protect\citename{Busso \bgroup et al.\egroup }2017]{msp-improv}
Busso, C., Parthasarathy, S., Burmania, A., AbdelWahab, M., Sadoughi, N., and
  Provost, E.~M.
\newblock (2017).
\newblock Msp-improv: An acted corpus of dyadic interactions to study emotion
  perception.
\newblock {\em IEEE Transactions on Affective Computing}, 8(1):67--80.

\bibitem[\protect\citename{Cao \bgroup et al.\egroup }2014]{crema-d}
Cao, H., Cooper, D., Keutmann, M., Gur, R., Nenkova, A., and Verma, R.
\newblock (2014).
\newblock Crema-d: Crowd-sourced emotional multimodal actors dataset.
\newblock {\em IEEE transactions on affective computing}, 5:377--390, 10.

\bibitem[\protect\citename{Cao \bgroup et al.\egroup }2015]{cao2015speaker}
Cao, H., Verma, R., and Nenkova, A.
\newblock (2015).
\newblock Speaker-sensitive emotion recognition via ranking: Studies on acted
  and spontaneous speech.
\newblock {\em Computer speech \& language}, 29(1):186--202.

\bibitem[\protect\citename{Chawla \bgroup et al.\egroup }2002]{chawla2002smote}
Chawla, N.~V., Bowyer, K.~W., Hall, L.~O., and Kegelmeyer, W.~P.
\newblock (2002).
\newblock Smote: synthetic minority over-sampling technique.
\newblock {\em Journal of artificial intelligence research}, 16:321--357.

\bibitem[\protect\citename{Costantini \bgroup et al.\egroup }2014]{emovo}
Costantini, G., Iaderola, I., Paoloni, A., and Todisco, M.
\newblock (2014).
\newblock Emovo corpus: an italian emotional speech database.
\newblock In {\em LREC}.

\bibitem[\protect\citename{Cui \bgroup et al.\egroup }2021]{emovie}
Cui, C., Ren, Y., Liu, J., Chen, F., Huang, R., Lei, M., and Zhao, Z.
\newblock (2021).
\newblock Emovie: A mandarin emotion speech dataset with a simple emotional
  text-to-speech model.

\bibitem[\protect\citename{Ekman}1992]{Ekman}
Ekman, P.
\newblock (1992).
\newblock An argument for basic emotions.
\newblock {\em Cognition \& Emotion}, 6:169--200.

\bibitem[\protect\citename{Eleftheriou \bgroup et al.\egroup }2021]{icnlsp}
Eleftheriou, S., Koromilas, P., and Giannakopoulos, T.
\newblock (2021).
\newblock Automatic assessment of speaking skills using aural and textual
  information.
\newblock In {\em Proceedings of The Fourth International Conference on Natural
  Language and Speech Processing (ICNLSP 2021)}, pages 166--175, Trento, Italy,
  12--13 November. Association for Computational Linguistics.

\bibitem[\protect\citename{Giannakopoulos}2015]{giannakopoulos2015pyaudioanalysis}
Giannakopoulos, T.
\newblock (2015).
\newblock pyaudioanalysis: An open-source python library for audio signal
  analysis.
\newblock {\em PloS one}, 10(12).

\bibitem[\protect\citename{Gloor \bgroup et al.\egroup
  }2019]{Gloor2019MeasuringAA}
Gloor, P.~A., Ara{\~n}o, K.~A., and Guerrazzi, E.
\newblock (2019).
\newblock Measuring audience and actor emotions at a theater play through
  automatic emotion recognition from face, speech, and body sensors.

\bibitem[\protect\citename{Gournay \bgroup et al.\egroup }2018]{cafe}
Gournay, P., Lahaie, O., and Lefebvre, R.
\newblock (2018).
\newblock A canadian french emotional speech dataset.
\newblock In {\em Proceedings of the 9th ACM Multimedia Systems Conference},
  MMSys '18, page 399–402, New York, NY, USA. Association for Computing
  Machinery.

\bibitem[\protect\citename{Ioffe and Szegedy}2015]{10.5555/3045118.3045167}
Ioffe, S. and Szegedy, C.
\newblock (2015).
\newblock Batch normalization: Accelerating deep network training by reducing
  internal covariate shift.
\newblock In {\em Proceedings of the 32nd International Conference on
  International Conference on Machine Learning - Volume 37}, ICML'15, page
  448–456. JMLR.org.

\bibitem[\protect\citename{Jackson and ul haq}2011]{savee}
Jackson, P. and ul~haq, S.
\newblock (2011).
\newblock Surrey audio-visual expressed emotion (savee) database, 04.

\bibitem[\protect\citename{Kadiri \bgroup et al.\egroup }2014]{ave}
Kadiri, S.~R., Gangamohan, P., Mittal, V.~K., and Yegnanarayana, B.
\newblock (2014).
\newblock Naturalistic audio-visual emotion database.
\newblock In {\em ICON}.

\bibitem[\protect\citename{Koromilas and
  Giannakopoulos}2021]{koromilas2021deep}
Koromilas, P. and Giannakopoulos, T.
\newblock (2021).
\newblock Deep multimodal emotion recognition on human speech: A review.
\newblock {\em Applied Sciences}, 11(17):7962.

\bibitem[\protect\citename{Kossaifi \bgroup et al.\egroup }2021]{sewa}
Kossaifi, J., Walecki, R., Panagakis, Y., Shen, J., Schmitt, M., Ringeval, F.,
  Han, J., Pandit, V., Toisoul, A., Schuller, B., and et~al.
\newblock (2021).
\newblock Sewa db: A rich database for audio-visual emotion and sentiment
  research in the wild.
\newblock {\em IEEE Transactions on Pattern Analysis and Machine Intelligence},
  43(3):1022–1040, Mar.

\bibitem[\protect\citename{Livingstone and Russo}2018]{ravdess}
Livingstone, S.~R. and Russo, F.~A.
\newblock (2018).
\newblock The ryerson audio-visual database of emotional speech and song
  (ravdess): A dynamic, multimodal set of facial and vocal expressions in north
  american english.
\newblock {\em PLOS ONE}, 13(5):1--35, 05.

\bibitem[\protect\citename{Ma \bgroup et al.\egroup }2020]{ma2020survey}
Ma, Y., Nguyen, K.~L., Xing, F.~Z., and Cambria, E.
\newblock (2020).
\newblock A survey on empathetic dialogue systems.
\newblock {\em Information Fusion}, 64:50--70.

\bibitem[\protect\citename{Martinez-Lucas \bgroup et al.\egroup }2020]{podcast}
Martinez-Lucas, L., Abdelwahab, M., and Busso, C.
\newblock (2020).
\newblock The msp-conversation corpus.
\newblock {\em Interspeech 2020}.

\bibitem[\protect\citename{Metallinou \bgroup et al.\egroup
  }2012]{metallinou2012context}
Metallinou, A., Wollmer, M., Katsamanis, A., Eyben, F., Schuller, B., and
  Narayanan, S.
\newblock (2012).
\newblock Context-sensitive learning for enhanced audiovisual emotion
  classification.
\newblock {\em IEEE Transactions on Affective Computing}, 3(2):184--198.

\bibitem[\protect\citename{Muljono \bgroup et al.\egroup }2019]{indonesian}
Muljono, Prasetya, M.~R., Harjoko, A., and Supriyanto, C.
\newblock (2019).
\newblock Speech emotion recognition of indonesian movie audio tracks based on
  mfcc and svm.
\newblock In {\em 2019 International Conference on contemporary Computing and
  Informatics (IC3I)}, pages 22--25.

\bibitem[\protect\citename{Ortony and Turner}1990]{psycology2}
Ortony, A. and Turner, T.
\newblock (1990).
\newblock What's basic about basic emotions?
\newblock {\em Psychological review}, 97:315--31, 08.

\bibitem[\protect\citename{Russell}1980]{circumplex}
Russell, J.
\newblock (1980).
\newblock A circumplex model of affect.
\newblock {\em Journal of Personality and Social Psychology}, 39:1161--1178,
  12.

\bibitem[\protect\citename{Vryzas \bgroup et al.\egroup }2018]{aesdd}
Vryzas, N., Kotsakis, R., Liatsou, A., Dimoulas, C., and Kalliris, G.
\newblock (2018).
\newblock Speech emotion recognition for performance interaction.
\newblock {\em Journal of the Audio Engineering Society. Audio Engineering
  Society}, 66:457--467, 06.

\bibitem[\protect\citename{Wundt and Judd}2011]{wundt2011outlines}
Wundt, W. and Judd, C.
\newblock (2011).
\newblock {\em Outlines of Psychology}.
\newblock W. Engelmann.

\end{thebibliography}

\label{lr:ref}
\bibliographystylelanguageresource{lrec2022-bib}

\end{document}